\documentclass{v18windows}

\def\pt{$p_{\mathrm{T}}$}

\def\be{\begin{equation}}
\def\ee{\end{equation}}
\def\bea{\begin{eqnarray}}
\def\eea{\end{eqnarray}}

\newcommand{\Photo}{\includegraphics[height=20mm]{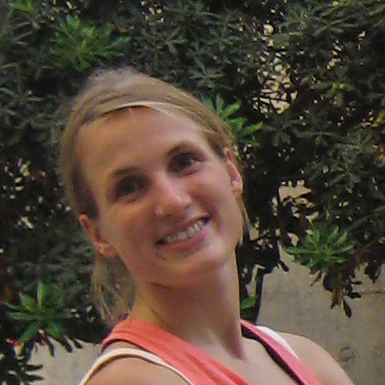}}

\begin{document}
\vspace*{4cm}
\title{SUSY HIGHLIGHTS -- CURRENT RESULTS AND FUTURE PROSPECTS}

\author{ Jory Sonneveld on behalf of the ATLAS and CMS Collaborations }

\address{Institute of Experimental Physics, Luruper Chaussee 149,\\
22761 Hamburg, Germany}

\maketitle\abstracts{
    At the Large Hadron Collider (LHC) at CERN in Geneva, Switzerland, one of the goals of colliding protons together is to search for new physics. Supersymmetry (SUSY), a popular theory of physics beyond the well-established standard model of particle physics, is a large part of the search program of the ATLAS and CMS multi-purpose detectors located on opposite sides of the 27-kilometer LHC ring.
    So far no sign of supersymmetry has been found in the most obvious search channels such as all-hadronic searches with jets and missing energy. This led to the development of many new search strategies. A selection of current results, including several observations of deviations from the standard model and novel search techniques, as well as future prospects are discussed in this talk.
}

\section{Introduction}
The Large Hadron Collider~\cite{Evans:2008zzb} at CERN in Geneva collides bunches of $10^{11}$ protons at 6.5 tera-electronvolt (TeV) per beam up to 1 billion times a second. This is the largest and most powerful collider in the world.
Seven detectors on the 27-kilometer ring are used to collect data from the 4 collision points on the LHC. Two multi-purpose detectors, A Toroidal LHC ApparatuS (ATLAS)~\cite{Aad:2008zzm} and a Compact Muon Solenoid (CMS)~\cite{Chatrchyan:2008aa}, on opposite sides of the LHC ring, are both competitive and complementary in measurements of the established standard model of particle physics (see e.g. an introduction by Aitchison~\cite{aitchison}
)
as well as in searches for new models of physics.

One of the many new models of physics is based on the only possible way to combine spacetime symmetries (the Poincar\'{e} group), such as translations and Lorentz transformations, with the internal symmetries SU(3)$\times$SU(2)$\times$U(1) that describe the standard model. In this combination, a new symmetry called supersymmetry associates bosons with fermions (for a review, see e.g. one by the particle data group~\cite{pdg18}).

At a hadron collider like the LHC, one can expect a large cross section for particles produced through the strong interaction, which in supersymmetry would be partners of the quarks and gluons called squarks and gluinos denoted as $\tilde{q}$ and $\tilde{g}$ (see Fig.~\ref{fig:xsecsres}) that result in multijet events. The supersymmetric electroweak gauge boson partners called gauginos and higgs boson partners called gauginos mix to states called neutralinos and charginos (also known as electroweakinos) denoted as $\tilde{\chi}^0$ and $\tilde{\chi}^\pm$. In a version of supersymmetry that, in order to constrain proton decay, is extended with a symmetry called R-parity, the lightest supersymmetry partners (LSPs), typically neutralinos, are stable, resulting in a lot of missing energy. This is typically expected from any dark matter candidate. This makes multijet and missing energy events an obvious signature for supersymmetry candidate events.

Electroweak production of supersymmetric particles has much lower cross sections than squark or gluino production. Electroweak supersymmetry partners typically decay to leptons and LSPs resulting in multilepton events with missing energy.

Some generic ways to look for new physics are resonance searches (also called ``bump hunts"), making use of angular distributions, and looking for deviations from standard model observables. The latter is not in the scope of this talk; here the focus is on direct searches for supersymmetry.

After several years of LHC collisions at 8 and 13 TeV, no obvious sign of supersymmetry has been observed. For this reason, many new search strategies were developed, among others techniques to search in events with very little missing energy that are characteristic of a variant of supersymmetry without light stable SUSY particles where R-parity is violated. Another challenging signature is that of very small mass (split) electroweakinos resulting in very soft leptons that are hard to trigger on and difficult to distinguish from background, if their tracks are reconstructed at all. In addition, more complex signatures such as those from longer lived particles and third-generation initial and final states now constitute a large part of the ATLAS and CMS supersymmetry search programs.

\section{Strong sparticle production}
\begin{figure}
\centering
\includegraphics[width=0.5\linewidth, trim=225 195 0 0, clip]{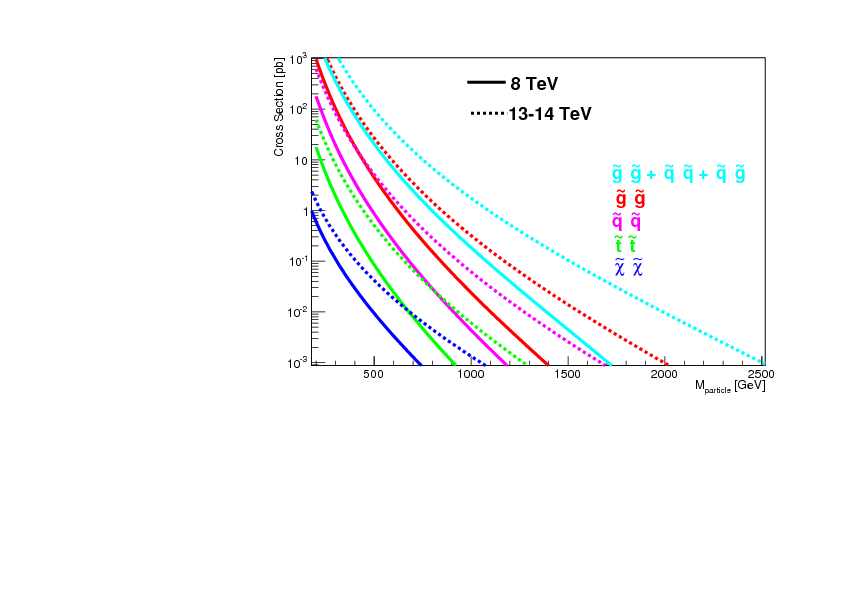}
\includegraphics[width=0.4\linewidth]{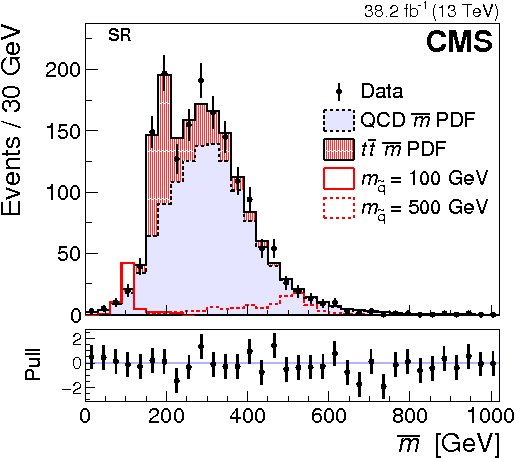}
    \caption[]{Left: Production cross sections of supersymmetric particles at the 13 TeV LHC. Electroweak sparticles such as neutralinos and charginos have much lower cross sections than the strongly produced gluinos and squarks~\cite{Ventura:2291346}.
    Right: CMS resonance search with expected signal from R-parity violating supersymmetry overlaid on the background~\cite{Sirunyan:2018zyx}.}
\label{fig:xsecsres}
\end{figure}

One obvious signature of supersymmetry is that of strong sparticle production, as this has the largest cross section at a hadron collider like the LHC; this is shown on the left in Fig.~\ref{fig:xsecsres}.
One way of searching for new physics is to search for so-called resonances, or ``bumps" in distributions of observables on top of standard model backgrounds, as shown on the right in Fig.~\ref{fig:xsecsres}. CMS conducted a resonance search for squark and gluino production in R-parity violating models of supersymmetry~\cite{Sirunyan:2018zyx}, where events with 4 to 5 quarks in the final state were selected with two jets with high \pt $ ~> 400$ GeV with substructure to suppress the QCD background that was estimated from data. This search placed the first constraints on pair-produced sparticles with masses below 400 GeV.

\begin{figure}
\centering
\centerline{\includegraphics[width=0.35\linewidth]{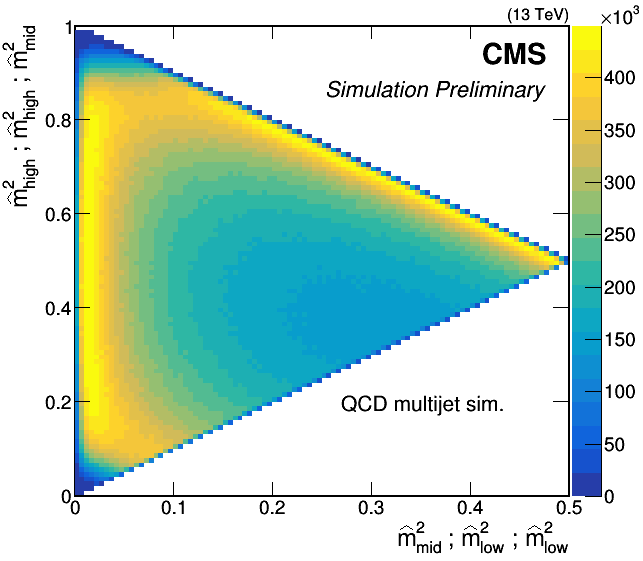}
\includegraphics[width=0.35\linewidth]{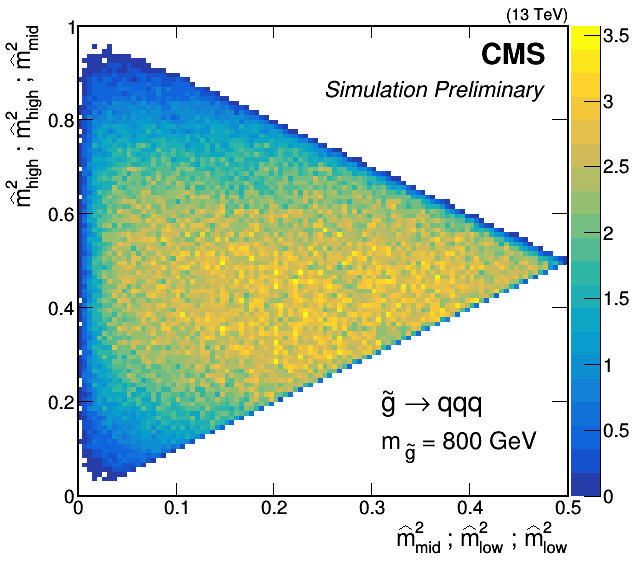}}
    \caption[]{Dalitz plots for QCD multijet background (left) and a signal of gluino pair production in R-parity violating supersymmetry with 6 jets in the final state in CMS data~\cite{CMS:2018ktt}.}
\label{fig:dalitz}
\end{figure}

Another resonance search for gluino pair production with R-parity violating decays in CMS data selected events with 6 jets in the final state, and made use of data scouting for low masses~\cite{CMS:2018ktt}. In data scouting, all events are saved, albeit in a trimmed format discarding some event information; in this case, a high level trigger selection saved all events at a 2kHz rate for a sum of jet transverse momenta $H_{\mathrm{T}} > 410$ GeV. From the resulting Dalitz plots shown in Fig.~\ref{fig:dalitz} one can see that the background can be well-separated from the signal.

One ATLAS search for strong production of SUSY particles explored the compressed supersymmetry mass region using charm tagging~\cite{Aaboud:2018zjf}, where they looked for stop and scharm production with decays to charms and neutralinos in events with at least three jets, where the nonleading jets were charm-tagged with an MVA that has a 18\% working point efficiency. Limits from this ATLAS search for stop and scharm production greatly improve those set by earlier ATLAS searches~\cite{Aaboud:2018zjf}.

\begin{figure}
\centering
\begin{minipage}{0.5\linewidth}
\includegraphics[width=\linewidth]{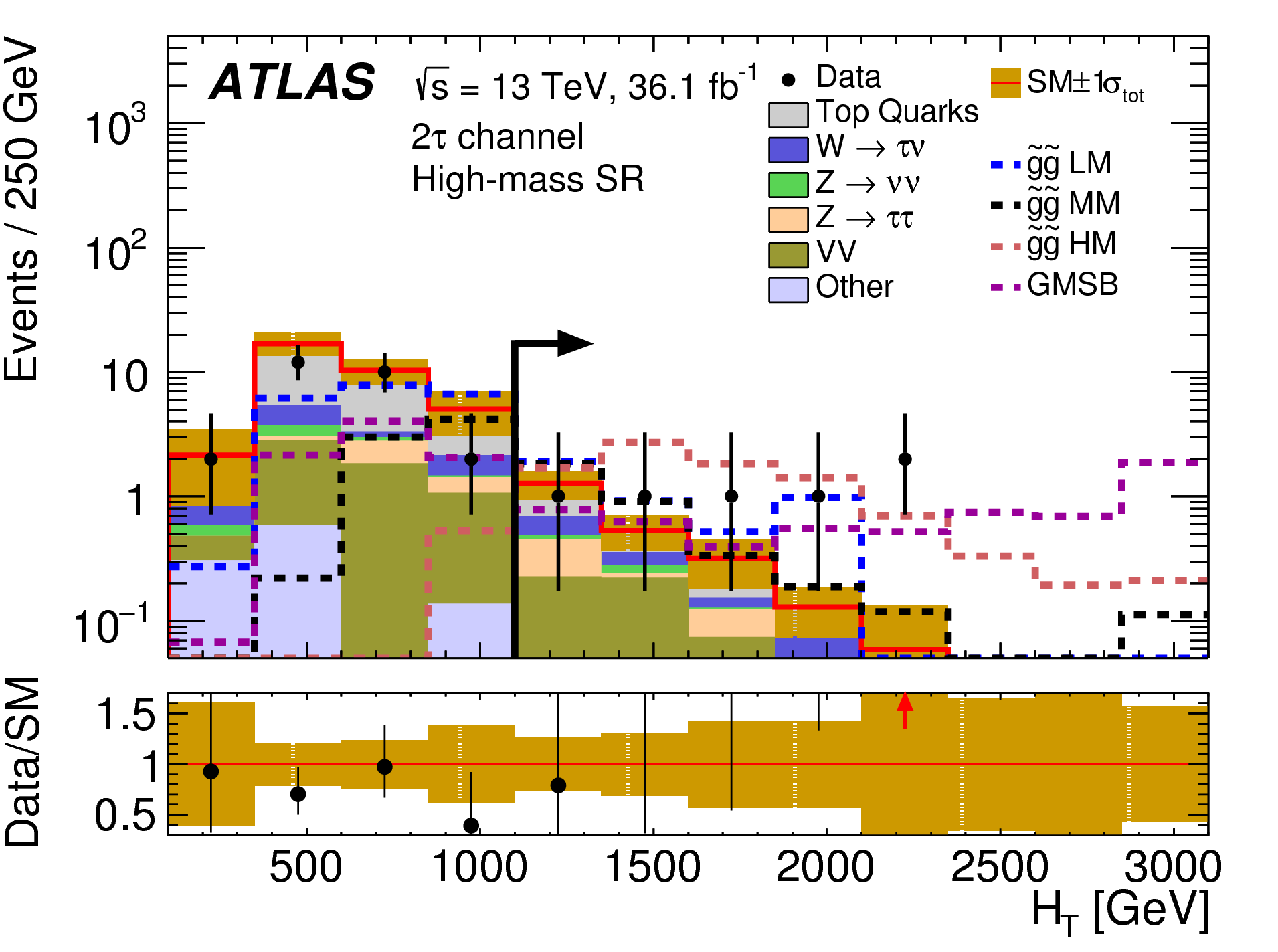}
\end{minipage}
\begin{minipage}{0.2\linewidth}
\includegraphics[width=\linewidth]{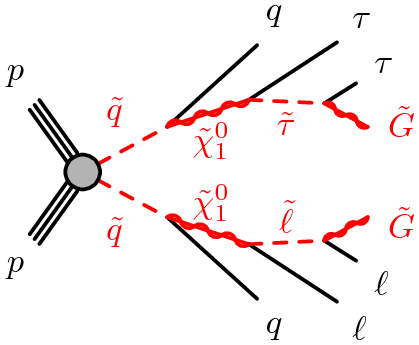}\\
\includegraphics[width=\linewidth]{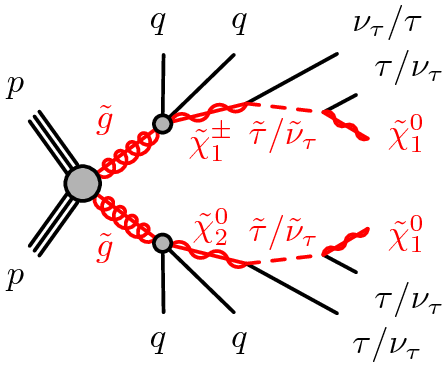}
\end{minipage}
    \caption[]{ATLAS search for squark and gluino production with taus in the final state. The expected signal for simplified models as well as a gauge-mediated supersymmetry breaking (GMSB) model are shown by the dotted lines~\cite{Aaboud:2018mna}.}
\label{fig:atlastaus}
\end{figure}

\begin{figure}
\centering
\centerline{\includegraphics[width=0.5\linewidth]{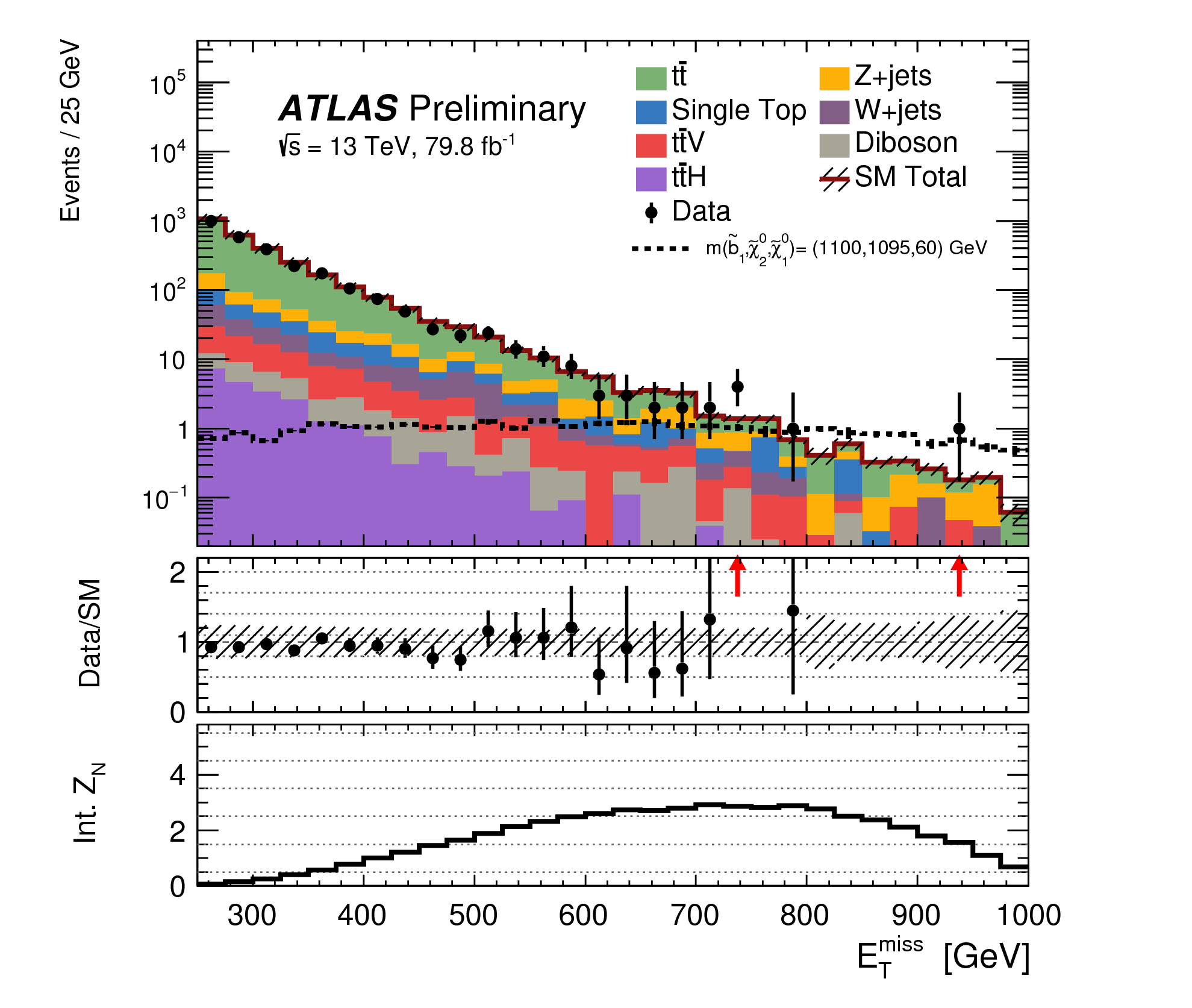}
\includegraphics[width=0.25\linewidth]{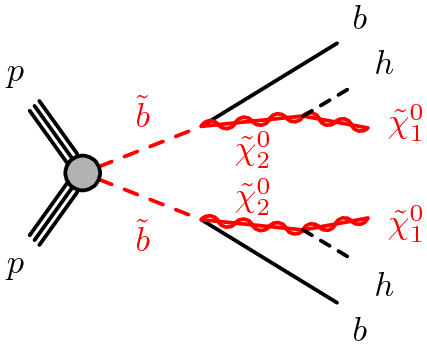}}
    \caption[]{ATLAS search for sbottom pair production with b quarks and higgs bosons in the final state. The expected signal for a simplified model is shown by the dotted lines~\cite{ATLAS:2018pnn}.}
\label{fig:atlassbottoms}
\end{figure}

Two strong sparticle production searches that saw an excess were an ATLAS search for tau leptons in the final state and and ATLAS search for sbottom pair production with b quarks in the final state. Searches with taus are not easy, as taus are often rejected with $t\bar{t}$ and $Wt$ backgrounds. This ATLAS search~\cite{Aaboud:2018mna} has a 40-60\% $\tau$ identification efficiency, and found a $2\sigma$ excess in events with one $\tau$ lepton and missing transverse momentum \pt $~> 400$ GeV or more than one tau lepton and \pt $~> 180$ GeV, as can be seen in Fig.~\ref{fig:atlastaus}.
A small excess was also observed in a search for sbottom pair production with the ATLAS detector~\cite{ATLAS:2018pnn} that decay as in Fig.~\ref{fig:atlassbottoms} in the compressed region with soft b-jets, as can be seen in Fig.~\ref{fig:atlassbottoms}.
A similar search was carried out in CMS~\cite{Sirunyan:2017eie}.

\section{Electroweak sparticle production}
\begin{figure}
\centerline{\includegraphics[width=0.6\linewidth]{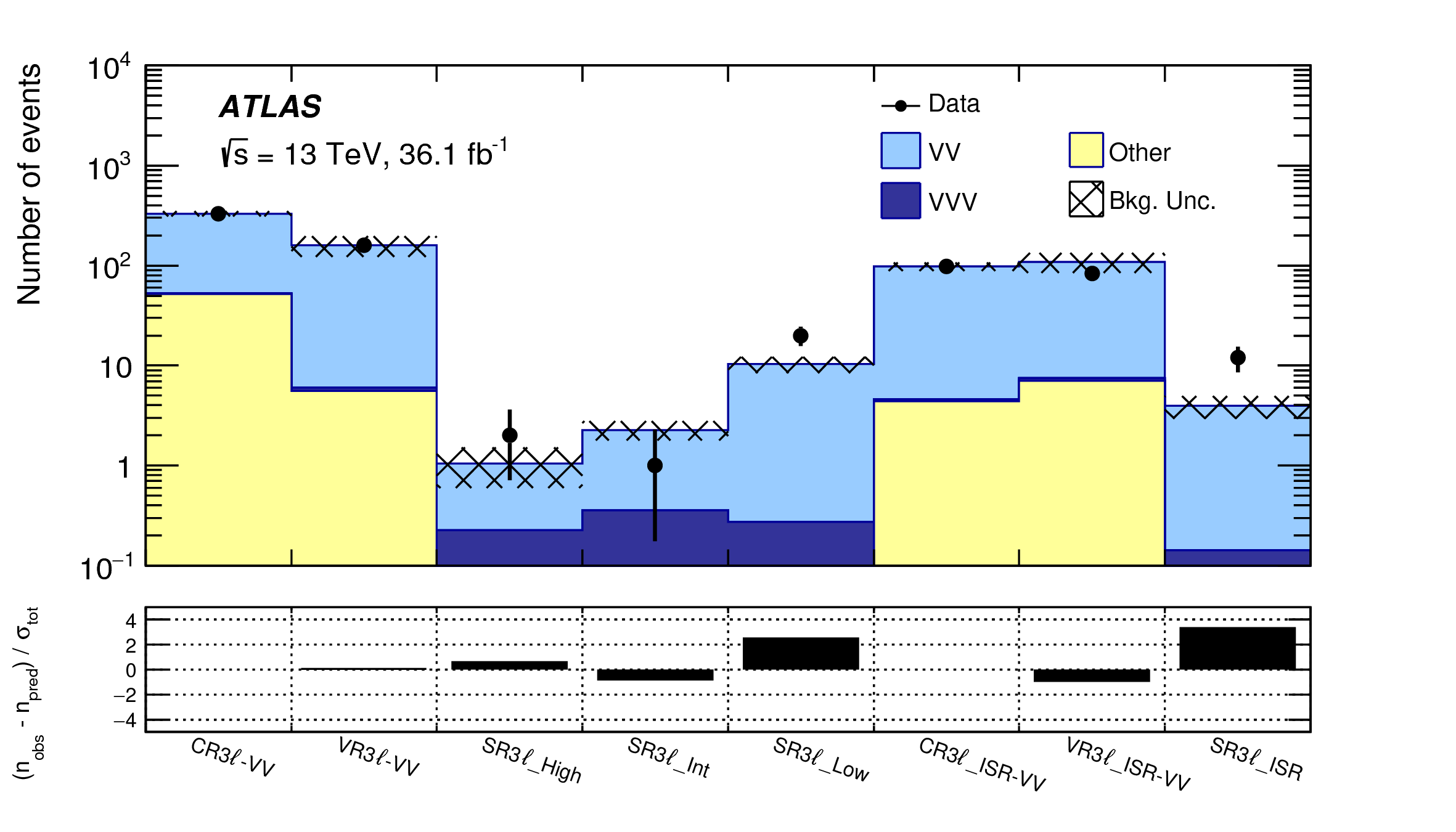}
\includegraphics[width=0.25\linewidth]{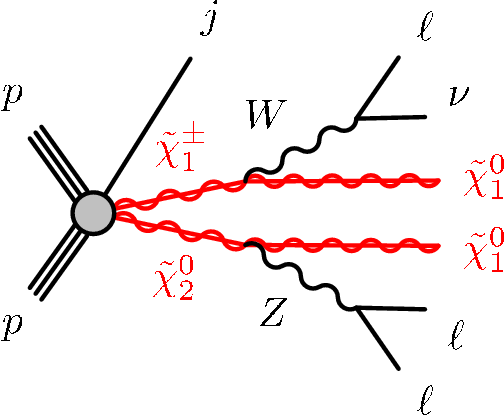}}
    \caption[]{ATLAS search for electroweakinos using a recursive jigsaw technique~\cite{Aaboud:2018sua}. A slight excess of $3\sigma$ is observed in the 3 lepton and ISR search region. The supersymmetry diagram corresponding to this signal is depicted on the right.}
\label{fig:recursive}
\end{figure}

Novel techniques were also employed in searches for electroweak sparticles. Classic $WZ$-mediated electroweakino searches are typically not sensitive to small mass splittings between charginos or second-lightest neutralinos and the lightest neutralino such as the CMS search for electroweakinos~\cite{Sirunyan:2018ubx}. For the process shown on the right in Fig.~\ref{fig:recursive}, a new technique called recursive jigsaw reconstruction was used. This technique uses kinematic observables computed in different reference frames to enhance sensitivity~\cite{Aaboud:2018sua}. An excess of $3\sigma$ was observed in one of the search regions as shown in Fig.~\ref{fig:recursive}.

If supersymmetry has a light higgsino, which is required by naturalness through the tendency to a lower $\mu$ parameter that determines the higgsino mass, and if the lightest neutralinos are higgsino-like, very soft leptons that lead to disappearing tracks are typical signatures. In dedicated searches for compressed electroweakino spectra neither ATLAS~\cite{Aaboud:2017leg} nor CMS~\cite{Sirunyan:2018iwl} observed anything beyond standard model backgrounds.
Searches for 4 leptons in the final state for both R-parity conserving and R-parity violating models can be found in both a CMS~\cite{CMS-PAS-SUS-17-002} and an ATLAS search~\cite{Aaboud:2018zeb}. Sensitivity is reduced for third-generation leptons, as is expected from tau reconstruction efficiencies.

\section{Longlived sparticles}
\begin{figure}
\includegraphics[width=0.4\linewidth]{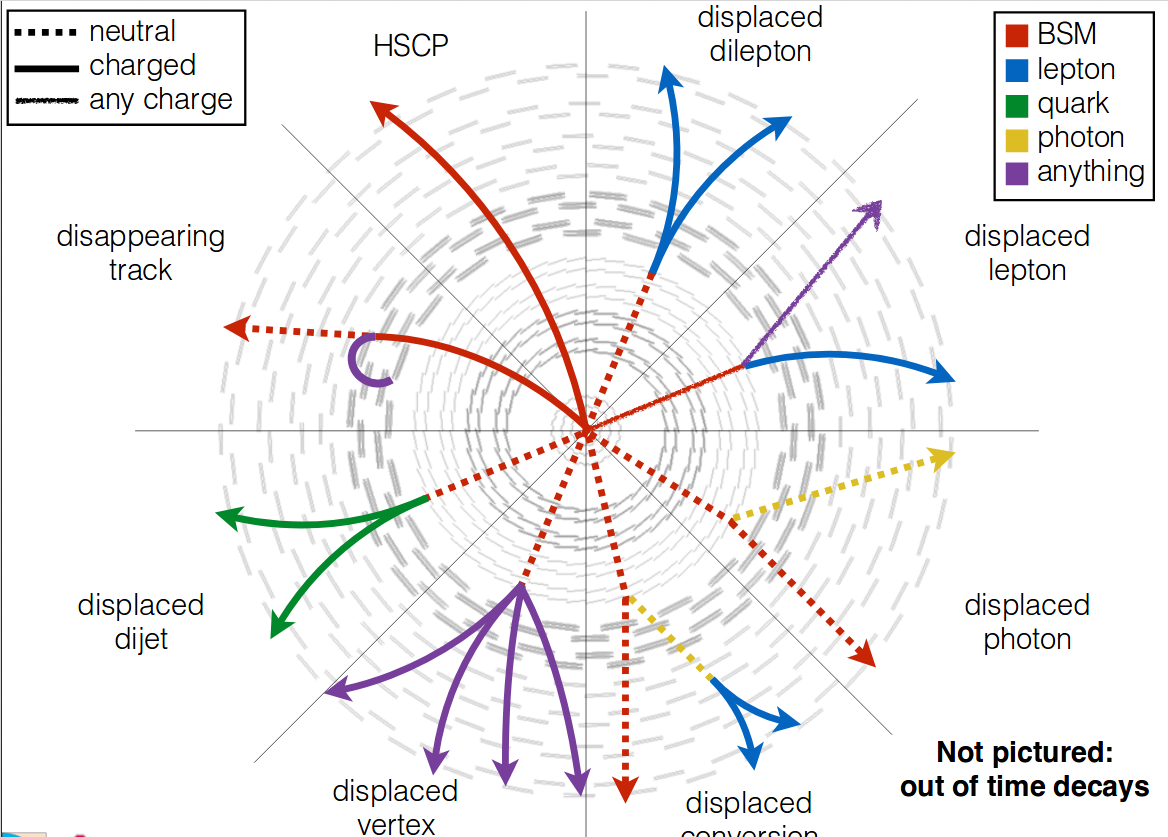}
\includegraphics[width=0.6\linewidth]{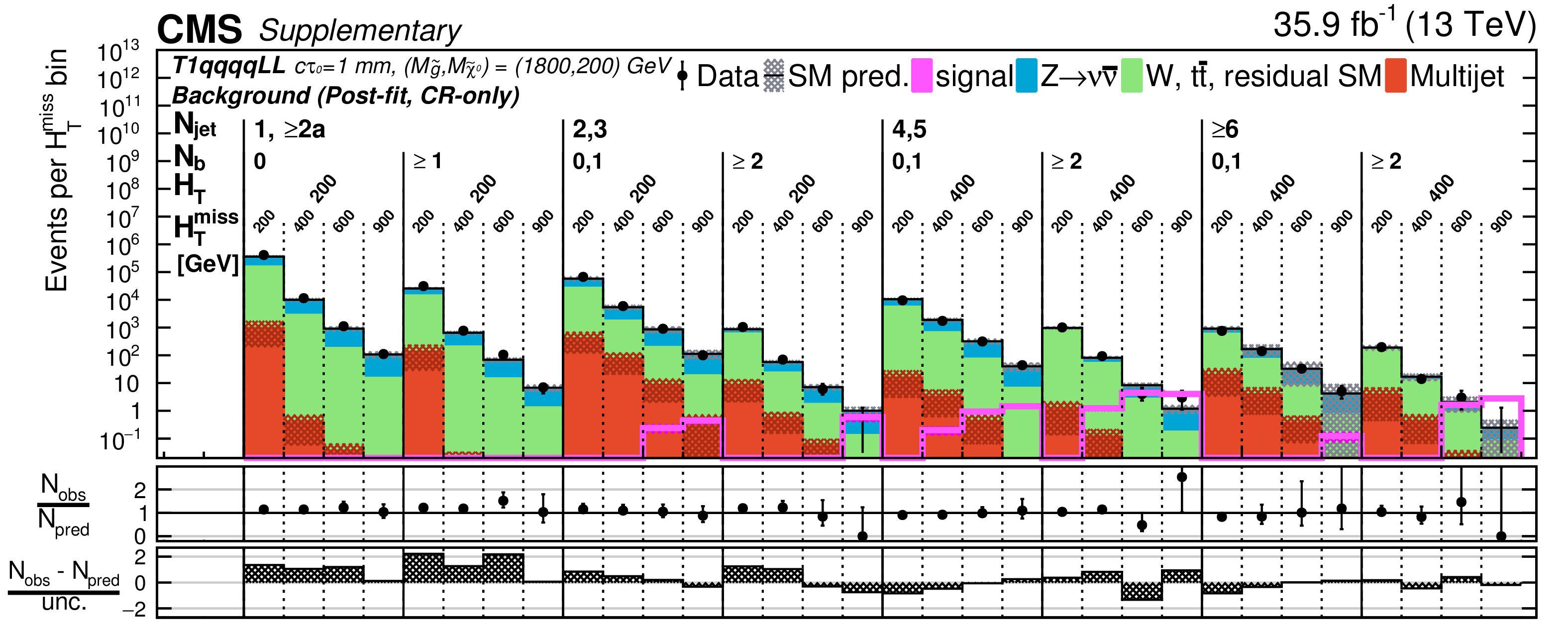}
    \caption[]{Left: Characteristic signatures of longlived particles in a multipurpose detector like ATLAS or CMS~\cite{antonelli}.
    Right: CMS inclusive search for supersymmetry in events with jets and missing energy with a longlived particle signal overlaid (right)~\cite{Sirunyan:2018vjp}.
}
\label{fig:cmslonglived}
\end{figure}

\begin{figure}
\centering
\includegraphics[width=0.45\linewidth]{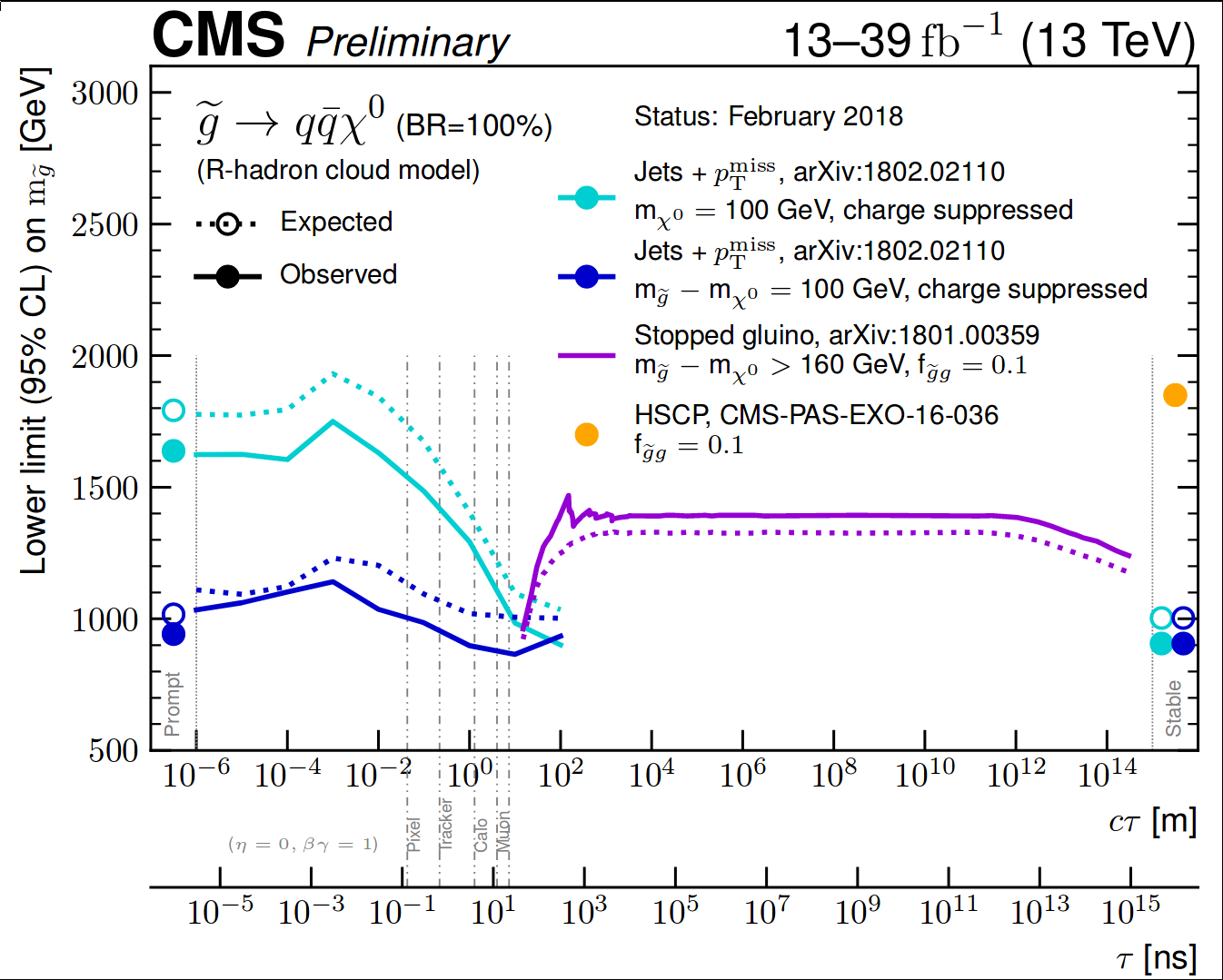}
\includegraphics[width=0.45\linewidth]{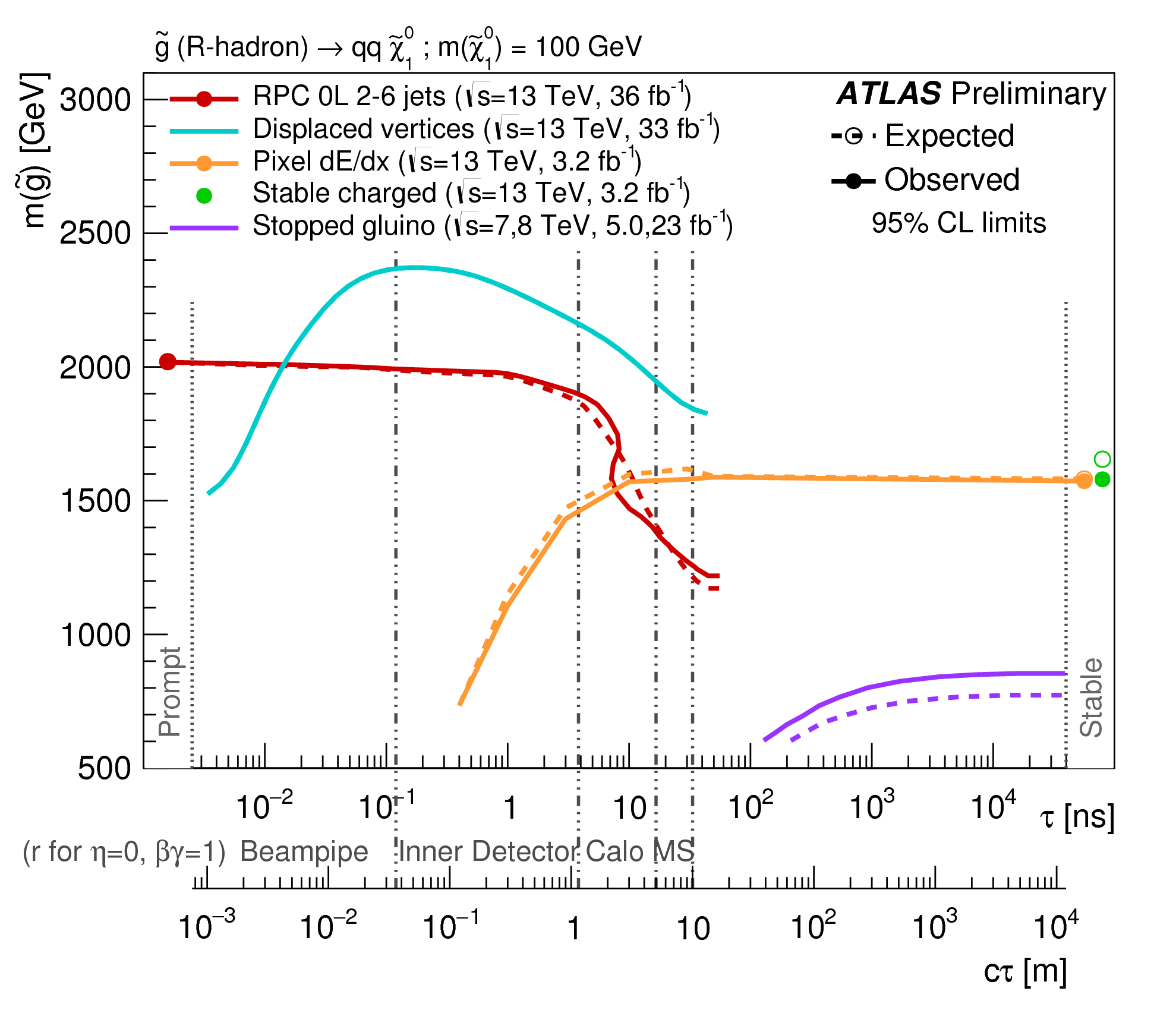}
    \caption[]{
Left:CMS inclusive search for supersymmetry in events with jets and missing energy, reinterpreted in terms of the R-hadron cloud model.
Results for prompt and stable gluinos are shown by the dots in the figure on the right~\cite{longlivedtwiki}.
Right: ATLAS inclusive search for supersymmetry in events with jets and missing energy (red) as well as targeted longlived particle searches~\cite{ATLAS:2018yey}.
Note how different searches are sensitive at different radii of the detector.
}
\label{fig:longlived}
\end{figure}

Many inclusive searches for conventional models of supersymmetry in events with jets and missing transverse momentum were conducted for promptly decaying sparticles, but no sign of such supersymmetry models has been found in LHC data. One search that uses the so-called $\alpha_{\mathrm{t}}$ variable has reinterpreted its results in terms of a split supersymmetry model with longlived sparticles~\cite{Sirunyan:2018vjp}. In this model, sfermions are light as well as the higgs, gluinos are assumed to be so-called R-hadrons, and the rest is heavy. Heavy quarks can make a gluino longlived; it becomes an R-hadron, or bound color-singlet state, that contains squarks or gluons and eventually decays to a quark, antiquark and LSP. Results for this model are shown in counts in Fig.~\ref{fig:cmslonglived} and in limits on lifetime and mass in Fig.~\ref{fig:longlived}. For comparison, another search for heavy stable charged particles (HSCP)~\cite{CMS:2016ybj} is shown. Such HSCP searches will greatly benefit from the tracker and muon phase~2 upgrades~\cite{tdr014,tdr016}.
 ATLAS, too, did such a reinterpretation in terms of R-hadrons~\cite{ATLAS:2018yey}, as shown on the right in Fig.~\ref{fig:longlived}.

Particles with longer lifetimes decay in characteristic ways in or outside our detector as shown on the left in Fig.~\ref{fig:cmslonglived}.
One such signature is a disappearing track, where often missing outer hits in the tracker and no calorimeter energy (to avoid leptons reconstructed as charged hadrons) are required. One example is a chargino that decays to a neutralino and a pion, where the pion is too soft to be reconstructed. Such searches were conducted with both ATLAS data~\cite{Aaboud:2017mpt} and with CMS data~\cite{Sirunyan:2018ldc}.

\begin{figure}
\centering
\begin{minipage}{0.37\linewidth}
\includegraphics[width=0.8\linewidth]{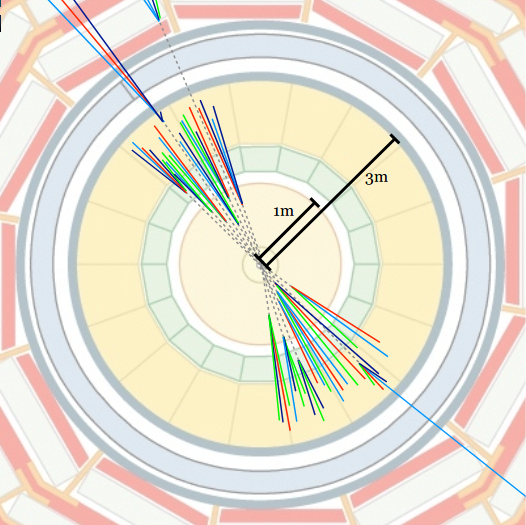}
\end{minipage}
\begin{minipage}{0.43\linewidth}
\includegraphics[width=\linewidth]{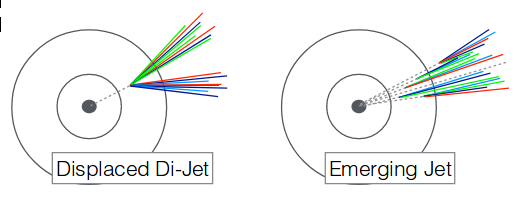}\\
\includegraphics[width=0.8\linewidth]{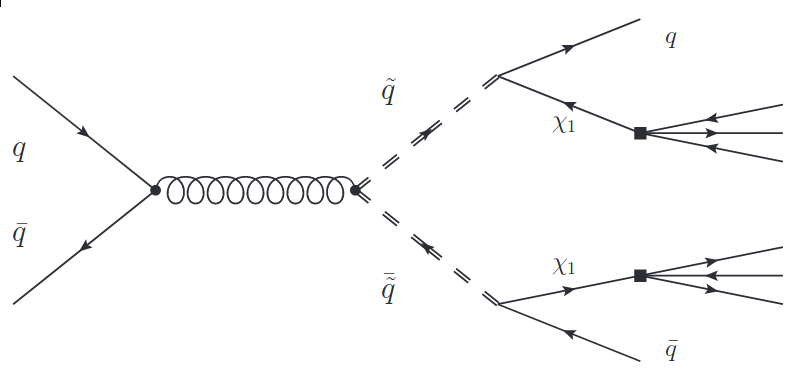}
\end{minipage}
    \caption[]{Signatures of emerging jets as they would look like at an LHC multipurpose detector (left). They are also shown as opposed to displaced jet signatures (middle). A corresponding diagram for emerging jets from a supersymmetric process is also shown (right)~\cite{Schwaller:2015gea}.}
\label{fig:emerging}
\end{figure}
\begin{figure}
\begin{minipage}{0.33\linewidth}
\includegraphics[width=\linewidth]{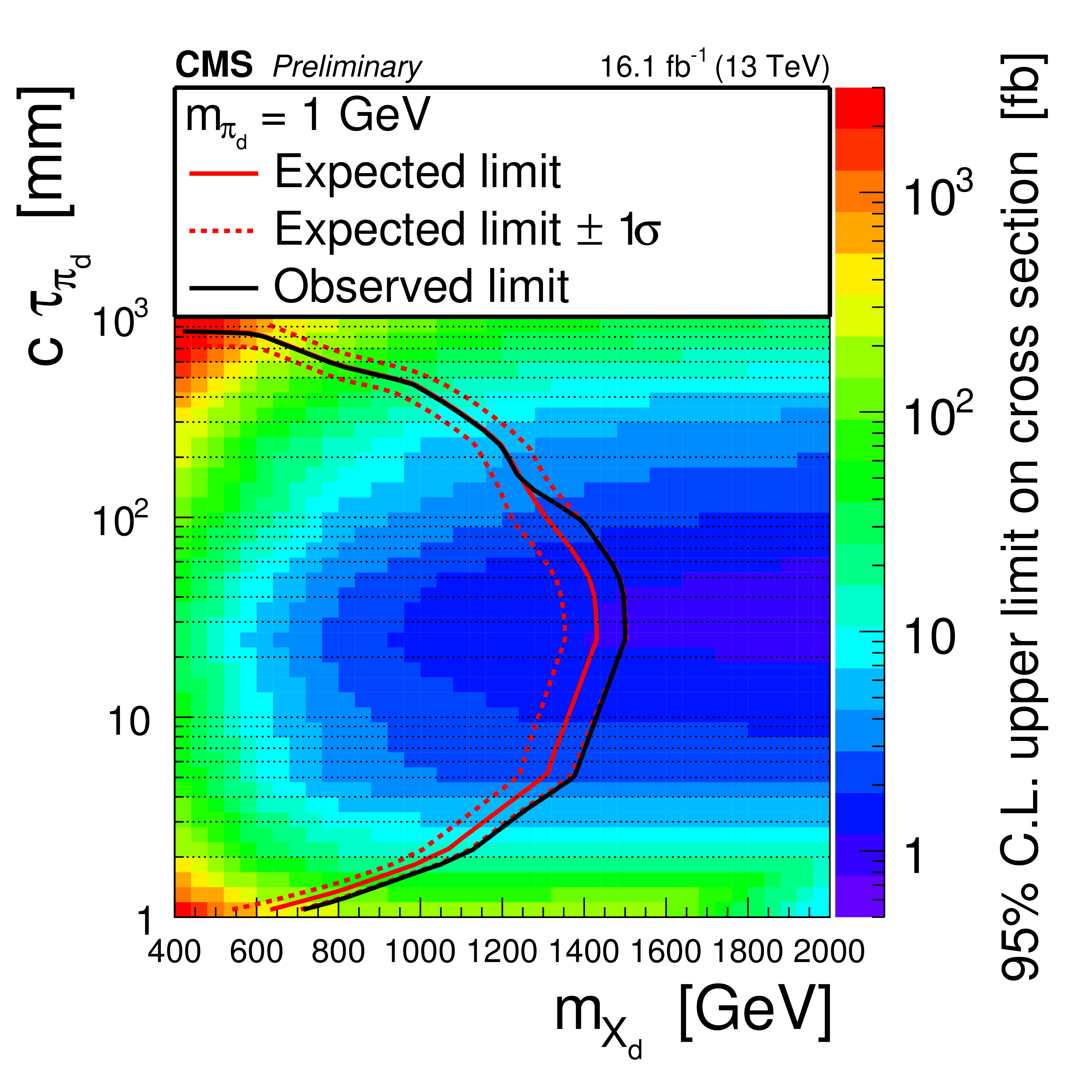}
\end{minipage}
\begin{minipage}{0.33\linewidth}
\includegraphics[width=\linewidth]{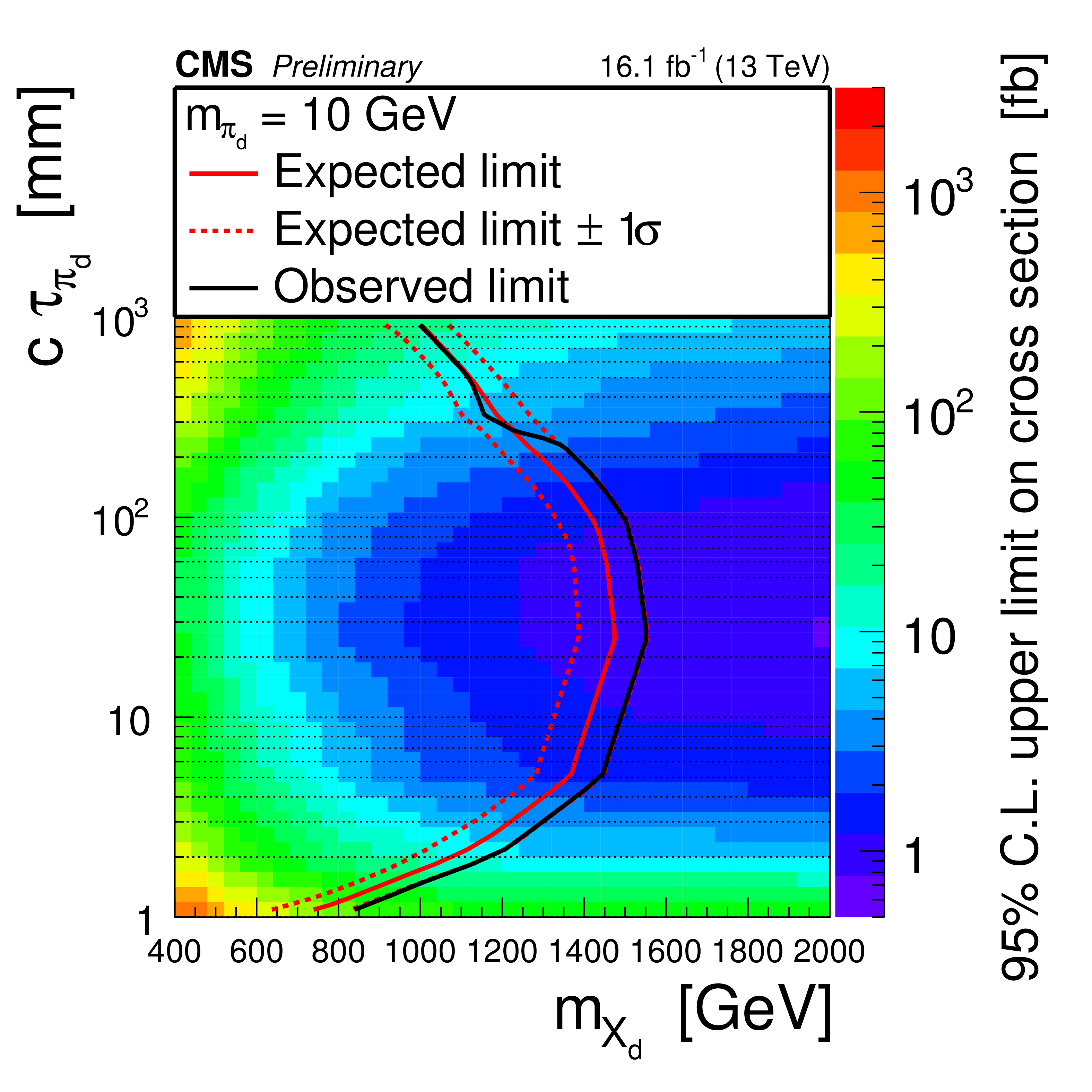}
\end{minipage}
\begin{minipage}{0.33\linewidth}
\includegraphics[width=0.7\linewidth]{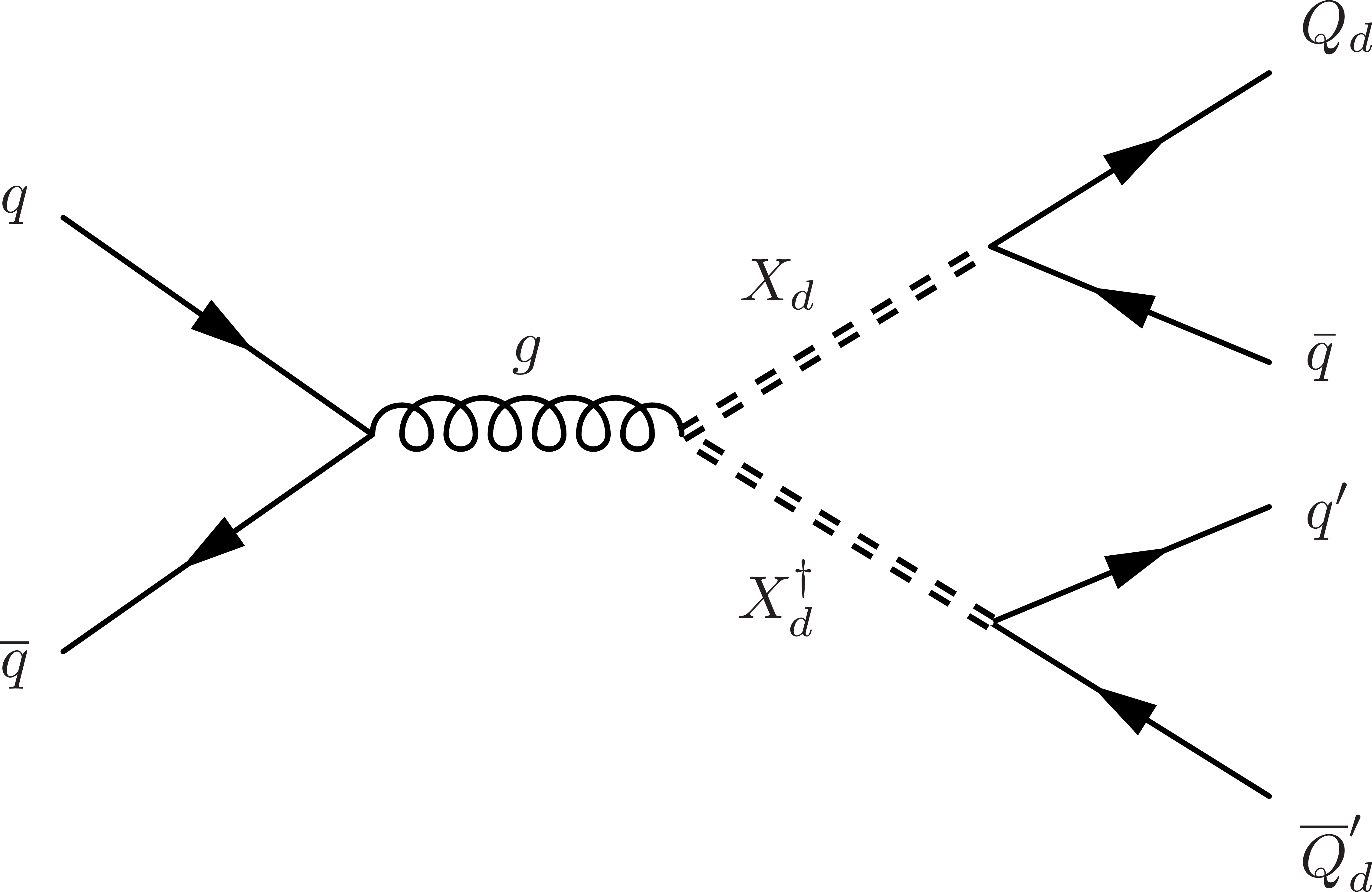}\\
\includegraphics[width=0.7\linewidth]{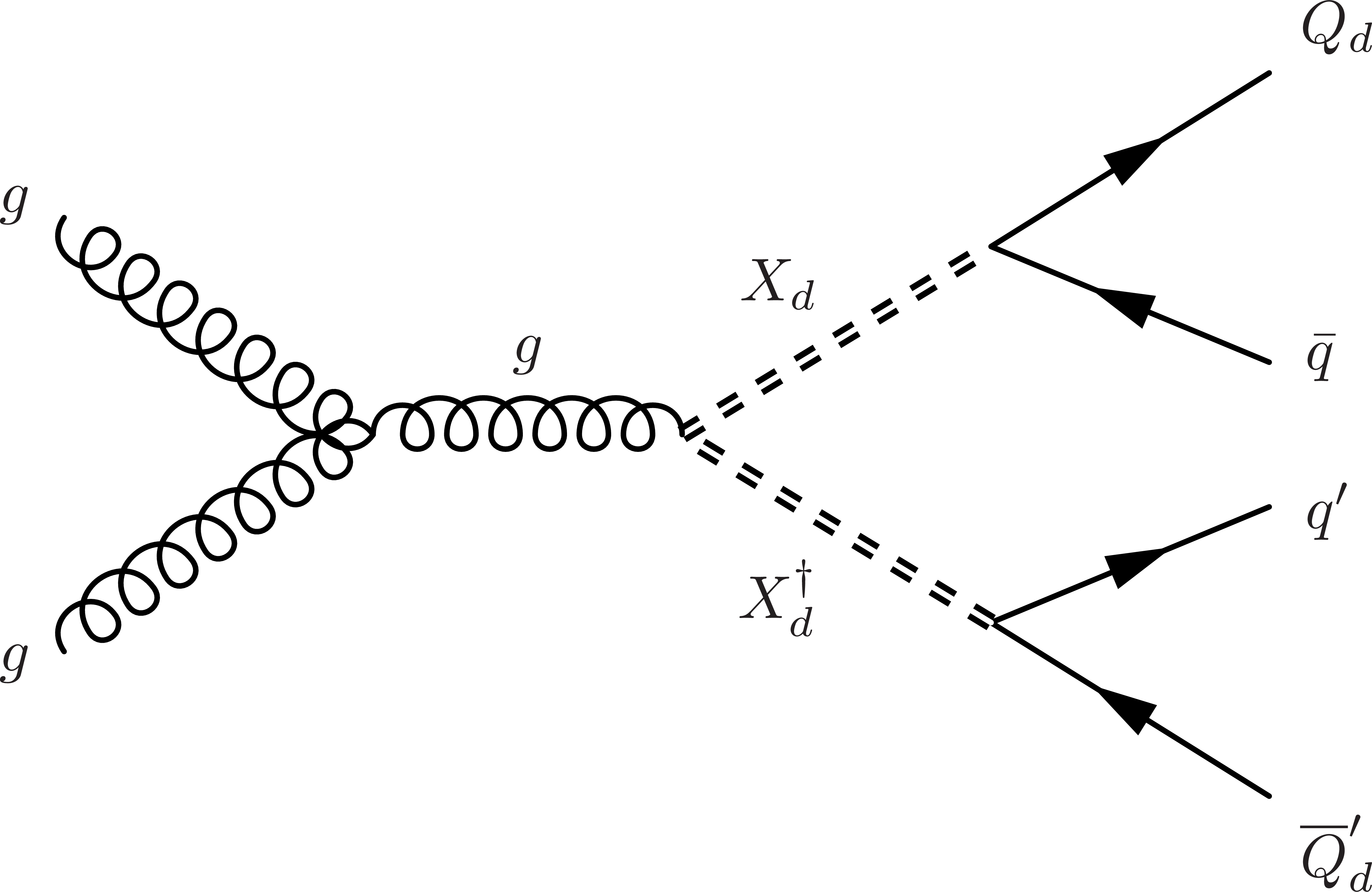}
\end{minipage}
    \caption[]{Results from a search for emerging jets in CMS (left) for dark particle decays (right)~\cite{Sirunyan:2018njd}.}
\label{fig:cmsemerging}
\end{figure}

A rather new idea of longlived particle signatures is that of emerging jets~\cite{Schwaller:2015gea}, as shown in Fig.~\ref{fig:emerging}. In supersymmetry, this could manifest itself in an R-parity violating model in events with squark pair production where a neutralino with a macroscopic lifetime decays to a $uds$ final state.
CMS searched for such particles and, in the absence of a signal, set limits on the invisible particle mass and lifetime~\cite{Sirunyan:2018njd}, as shown in Fig.~\ref{fig:cmsemerging}.

\begin{figure}
\centering
\includegraphics[width=0.45\linewidth]{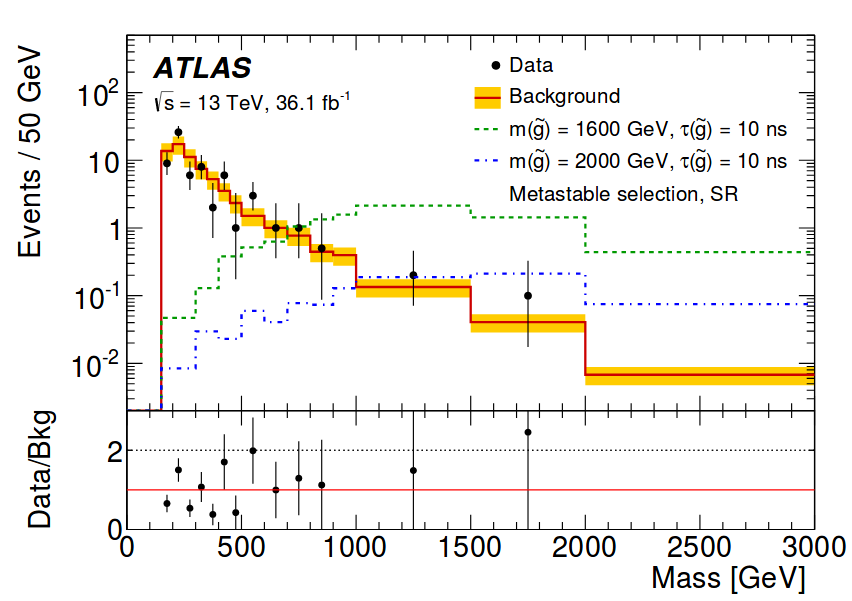}
\includegraphics[width=0.45\linewidth, trim=250 190 50 50, clip]{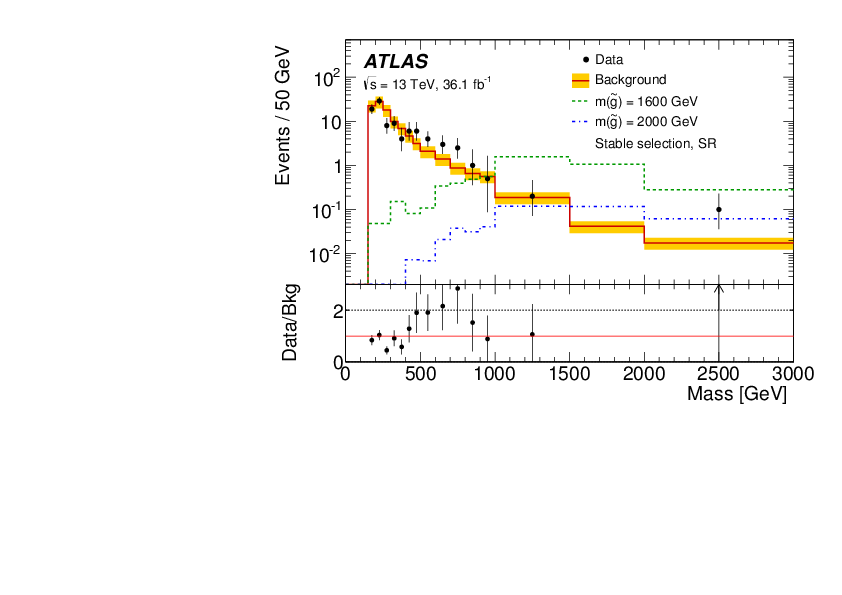}
    \caption[]{Results for both metastable (left) and stable (right) charged particles using an ionization measurement in the ATLAS detector. A slight excess of 2.4$\sigma$ has been observed in the results for the stable particle)~\cite{Aaboud:2018hdl}.}
\label{fig:atlasdedx}
\end{figure}

ATLAS performed two searches for more stable charged particles that had a longer reach in the detector, as was already shown in Fig.~\ref{fig:longlived}. One of these looked for stopped gluinos~\cite{Aad:2013gva} and the other used pixel dE/dx~\cite{Aaboud:2018hdl} for both stable particles that decay outside the ATLAS detector using ISR for triggering, and metastable charged particles that do not reach the muon spectrometer and live several to several tens of nanoseconds. Results are shown in Fig.~\ref{fig:atlasdedx}.
For this search, corrections to the measured dE/dx were made in (1) the most probable value (MPV) for the dE/dx distribution that changes with radiation damage, (2) low-momentum kaons and protons as a result of multiple scattering, (3) the traversed thickness, and (4) corrections to simulation to account for radiation damage in pixel sensors. In addition, only events with particles with $0.3 < \beta\gamma < 0.9$ were used, as for $\beta\gamma < 0.3$ the time over threshold is too long, and for $\beta\gamma > 0.9$ the background from standard model activity is too large.

\section{Beyond LHC run 2}
\begin{figure}
\centering
\includegraphics[width=0.55\linewidth]{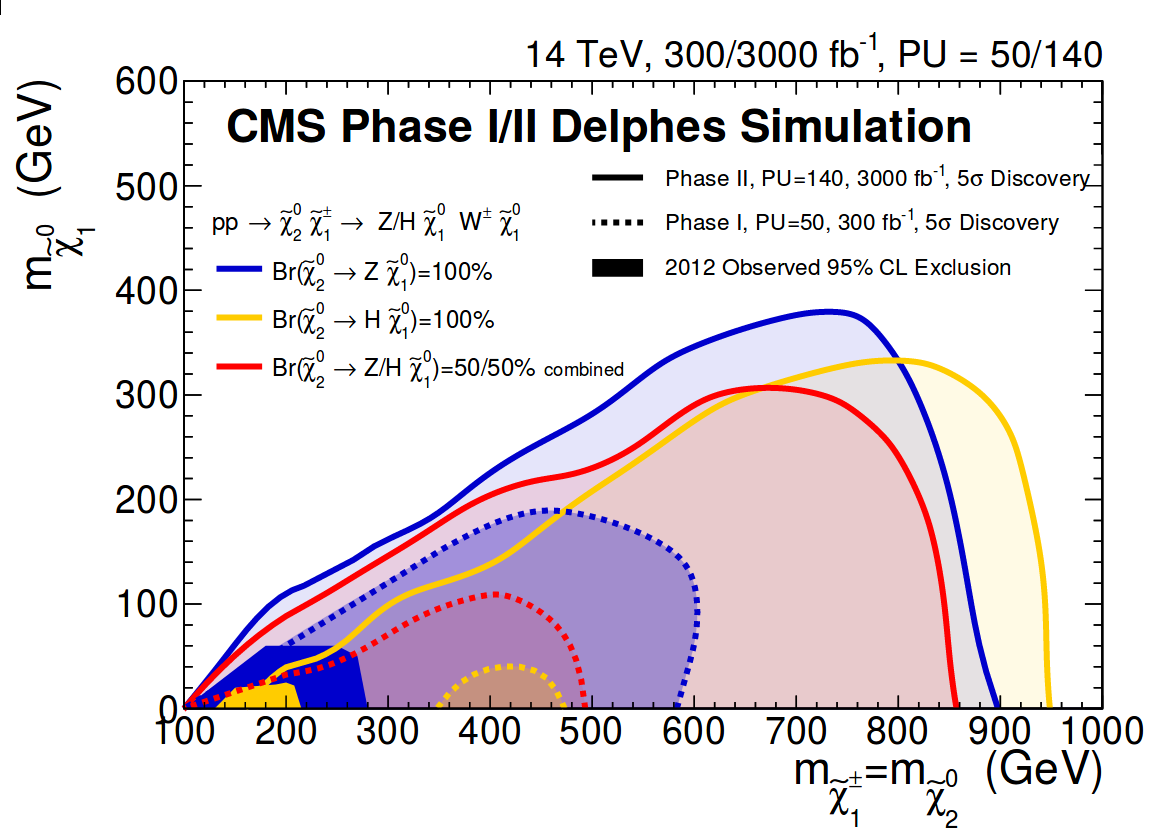}
    \caption[]{Expected improvement in limits on charginos and neutralinos in run 3 with data from the CMS detector~\cite{CMSCollaboration:2015zni}.}
\label{fig:cmsrun3}
\end{figure}

Proton-proton collisions in LHC run 2 have now come to an end. After a long shutdown of two years, run 3 will start, and in several years time, now planned in 2026, the high-luminosity LHC will start, where the level-1 trigger rate in CMS will increase from 100kHz to 750kHz, and in ATLAS the corresponding level-0 trigger rate increases to 1MHz.
In run 2, ATLAS and CMS have each recorded (received) 149 (158) and 151 (163) $\mathrm{fb}^{-1}$~\cite{atlaslumi,cmslumi}.
It is expected that in run 3 another 300 $\mathrm{fb}^{-1}$ will be collected, and at the high-luminosity LHC another 3000 $\mathrm{fb}^{-1}$ will be collected~\cite{lhcplans}, which will facilitate searches for larger mass sparticles as well as lower cross section models, where limits are expected to greatly improve as shown in Fig.~\ref{fig:cmsrun3}~\cite{CMSCollaboration:2015zni}.
On the other hand, pileup will reach 140, which will mean degradation in trigger efficiency, b-tagging, and resolution in missing transverse energy.

A small mass splitting between charginos and neutralinos will result in longlived charginos. The CMS phase~2 tracker will have a flag for ionizing particles in the outer tracker and excellent dE/dx resolution in the inner tracker~\cite{Klein:2017nke}, and the CMS muon time of flight system can be used to distinguish heavy stable charged particles from muons~\cite{tdr16}.
The ATLAS transition radiation tracker will be replaced with an all-silicon tracker that is expected to have a better momentum resolution with a larger outer radius than the current tracker \cite{atlasletterintent,ATLAS:2013hta}.
Similarly, the CMS phase~2 tracker will have an improved mass resolution that will make $B_s \to \mu\mu$ more easily distinguishable from $B_0 \to \mu\mu$~\cite{Klein:2017nke}.

\section{Summary and outlook}
Data taking at the LHC was successful at both ATLAS and CMS in the past years. Many new search techniques have been explored and new searches for models of supersymmetry with more complicated signatures have been carried out. Several excesses were observed, but there is still no clear sign of new physics.
In the LHC run 3, more statistics can lead to better sensitivity in searches for models with low cross sections like electroweak supersymmetry searches, and at the high-luminosity LHC track-triggers can help look for more unconventional signatures.
In addition, a well-advanced $e^+e^-$ collider program also offers good perspectives for more searches for supersymmetry beyond the LHC.

\end{document}